# Electron tunneling in chaotic InAs/GaAs quantum ring

I. Filikhin, S. G. Matinyan and B. Vlahovic

North Carolina Central University, Durham, NC 27707, USA

**Abstract**
Two dimensional InAs/GaAs quantum ring (QR) is considered using the effective potential approach. The symmetry of QR shape is violated as it is in the well-known Bohigas annular billiard. We calculate energy spectrum and studied the spatial localization of a single electron in such QR. For weak violation of the QR shape symmetry, the spectrum is presented as a set of quasi-doublets. Tunneling between quasi-doublet states is studied by the dependence on energy of the states. The dependence is changed with variation of the QR geometry that is related to the eccentricity of the QR. An interpretation of the experimental result obtained in [1] is proposed. We show that the "chaos-assisted tunneling" effect found in this paper can be explained by inter-band interactions occurred by anti-crossing of the levels with different "radial" quantum numbers.

**Introduction**

Tunneling and Chaos are inalienable phenomena in the meso- and nano world .Technology itself with its imperfection of shapes of Quantum Dots (QDs) and Quantum Rings (QRs) provides the chaotic behavior in the QD and QR [2] what has a strong influence on their transport and other properties.
   The problem of the quantum chaos in these objects has a relatively long history, just since they entered science and technology [3-7] (for recent review see [8]).
   One of the main results of these studies is that their chaotic phenomena sensitively depend on geometry and, first of all, on the symmetry of these objects.
   The second fundamental phenomenon characterizing the behavior of these confined objects is a tunneling expressing the matter wave character of Quantum Mechanics. The tunneling is well known from textbooks and best discussed by an example of the barrier penetration in a double well potential. However, in the general sense, as also applied to the confined nano objects, tunneling is a dynamical in essence [9], not necessarily associated with existing the energetic barrier, it is the property of the wave function of the whole system. The main feature of this so called dynamical tunneling is well known from energetic barrier examples splitting of the degenerate pairs of level when the coupling between objects due to their common wave function resulting the formation two nearly degenerate eigenstates which are the linear combinations of the wave functions of the isolated objects. In principle, this splitting $\Delta E$ is exponentially small [10] in analogy with the above energy barrier example.

Chaos introduces novel phenomenon in this problem [11]: it enhances the tunneling rate (energy average tunneling probability) between regular (non-chaotic) islands in the phase space. One can guess that the tunneling rate is determined not by the wave function overlap between the regular islands but rather by the overlap of the regular islands with the chaotic "sea" in two step way (see below). Then one finds an essential enhancement of the energy splitting $\Delta E$ and we arrive to the new type of interconnection between the chaos and tunneling - so called chaos assisted tunneling (CAT) [11-15].

We need to remark that the arguments for CAT use the phase space treatment (stroboscopic presentation). From the energetic and configuration picture using the semi classical approach CAT may be described as follows: It is known that the main behavior of the chaotic systems is regulated by periodic (unstable) orbits (PO) [16-18] and the spectrum of eigenvalues of Quantum Hamiltonian is dual to the spectrum of PO of the system. Moreover, the number of PO of chaotic system is exponentially large in comparison with the regular system with the power like amount of PO. One may guess that the picture of two step way of CAT proposed and developed in [14, 15] corresponds just to this exponential number of PO for chaotic system and leads to much higher splitting $\Delta E$ and thus to enhance the matrix element of the tunneling. In this picture the enhancement of the tunneling may have a resonance character due to the avoided crossings of the tunneling doublet eigenenergies with the eigenenergy of a state located in the chaotic phase space region (for details see book [19]).

The study of CAT includes among the others the remarkable Bohigas annular billiard [20] (see also [14], [15]). In this connection we need to stress that the studies of CAT as a rule even experimentally [1], [21] were provided for billiard like configurations. We consider here two dimensional (2D) chaotic quantum ring. Difference between the treatment of the Bohigas annular billiard and this QR is in the boundary conditions: the Dirichlet in the first case and the BenDaniel-Duke in the second. We consider the more realistic situation with a finite potential, and this is one of the goals of the present study of the above mentioned important phenomena as applied to the system of QDs and QRs.

**Description of the model**

We consider the InAs/GaAs QR with the shape of the annular billiard proposed by Bohigas et al. [20] and filled the area between two non-concentric circles of radii $R$ and $r$ centered at $(x, y)$ coordinates $(0,0)$ and $(0,-\delta)$. The QR is symmetric under reflections with respect to $y$-axis and asymmetric relative to $x$-axis.

These heterostructures are modelled utilizing the *kp*-perturbation single sub-band approach with an energy dependent quasi-particle effective mass. The problem is mathematically formulated by the Schrödinger equation in two dimensions:

$$(H_0 + V(\mathbf{r}) - E)\Psi(\mathbf{r}) = 0. \qquad (1)$$

Here $H_0$ is the single band **kp**-Hamiltonian operator $H_0 = -\nabla \frac{\hbar^2}{2m^*} \nabla$, $m^* = m^*(\mathbf{r}, E)$ is the electron effective mass which depends on both the energy and the position of the electron. $V(\mathbf{r})$ is confinement potential: $V(\mathbf{r}) = V_c(\mathbf{r}) + V_s(\mathbf{r})$, where $V_c(\mathbf{r})$ is the band gap potential. $V_c(\mathbf{r}) = 0$ inside the QR, and is equal to $V_c$ outside the QR, where $V_c$ is defined by the conduction band offset for the bulk.

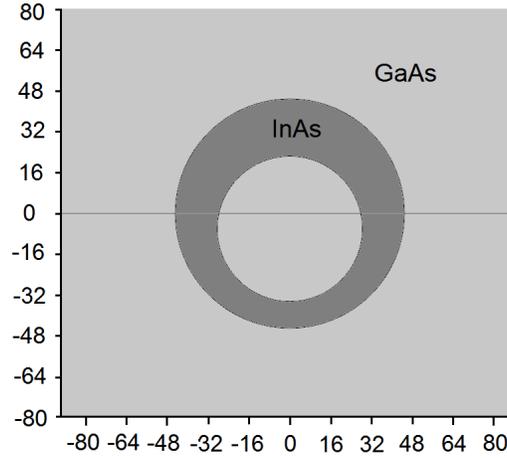

Figure 1. InAs/GaAs quantum ring. The sizes are given in nm. The horizontal line passes through the center of the outer circle.

The band gap potential for the conduction band was chosen as $V_c$ =0.594 eV. Bulk effective masses of InAs and GaAs are $m^*_{0,1}$=0.024 $m_0$ and $m^*_{0,2}$=0.067 $m_0$, respectively, where $m_0$ is the free electron mass. $V_s(\mathbf{r})$ is the effective potential simulating the strain effect [22]. The effective potential $V_s(\mathbf{r})$ has an attractive character and acts inside the volume of the QR. The magnitude of the potential can be chosen to reproduce experimental data. For example, the magnitude of $V_s$ for the conduction band chosen in [23] is 0.21 eV. This value was obtained to reproduce results of the 8-th band **kp**-calculations of Ref [24] for InAs/GaAs QD. To reproduce the experimental data from [25], the $V_s$ value of 0.31 eV was used in [22] for the conduction band. The difference of these values can be explained by inter-material mixing in experiment. The BenDaniel-Duke boundary conditions are used on the interface of the material of QR and substrate.

We can separate an angle variable in Eq. (1) for the QR shape having rotation symmetry when the centers of inner and outer circles coincide. In polar coordinates $(\rho,\varphi)$ the wave function is represented as

$$\Psi_{n,l}(\rho,\varphi) = \Phi_{n,l}(\rho)e^{il\varphi}, \qquad (2)$$

where $n = 1,2,3...$ and $l = \pm 0, \pm 1, \pm 2,...$ are orbital quantum numbers. In the case of violation of the shape symmetry one cannot definite "good" quantum numbers $(n,l)$. Followed by [20], we assume that there is the set of quantum number $(n,m)$ which is generalized analogue of radial and orbital quantum numbers $(n,l)$ used for regular quantum ring ($\delta$ =0). Let us formulate the condition for the approximation. The Eq. (1) for $\delta$ >0 can be rewritten as following

$$(H_0 + V_{sym}(\mathbf{r}) + V_{asym}(\mathbf{r}) - E)\Psi(\mathbf{r}) = 0.$$

Here $V_{sym}(\mathbf{r})$ and $V_{asym}(\mathbf{r})$ are defined by spatial regions $\Gamma_{sym}$ and $\Gamma_{asym}$ as it is shown in Fig. 2.

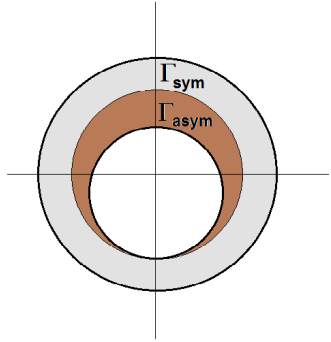

Figure 2. The spatial regions $\Gamma_{sym}$ and $\Gamma_{asym}$ are dividing the QR shape (see Fig.1) to the symmetrical and asymmetrical parts.

When the region $\Gamma_{sym}$ is dominating over $\Gamma_{asym}$, $V_{asym}(\mathbf{r})$ can be considered as a small perturbation of the confinement potential $V_{sym}(\mathbf{r})$ and the approximation with quantum numbers $(n,m)$ may be acceptable. Note that the part of Hamiltonian corresponding to $V_{sym}(\mathbf{r})$ produce centrifugal potential $\dfrac{l^2}{\rho}$ which will dominate for large $|l|$ over $V_{asym}(\mathbf{r})$ due to the large repulsive strength for short distances from the QR center. Formally, the $V_{asym}(\mathbf{r})$ may be also neglected for $|l|\to\infty$. However, the nanosized QR confinement has limitation for values of $|l|$.

The important geometry parameters are $d = R-(r+\delta)$ - width of the ring defined by the $\Gamma_{sym}$ area and $S = (r+\delta)/R$ - the value corresponding to the classical impact parameter [20] with its quantum counterpart $S = m/k/R$, where $m$ is angular momentum quantum number, $k$ is the wave number. For small $\delta$ the $d$ defines a maximal radial quantum number of the confinement states by the approximate relation $n_{max}^2 \leq V_c d^2$. The value $r+\delta$ gives the radius of the inner circle corresponding to the $\Gamma_{sym}$ area.

To describe tunneling of a single electron in this chaotic quantum object we have to make some definition. Probability of localization of electron in the region $\Omega_\gamma$ ($\gamma=1, 2$) is defined by $N_{n,\gamma} = \iint\limits_{\Omega_\gamma} |\Phi_n(x,y)|^2\, dxdy$, where $\Phi_n(x,y)$ is wave function of electron, $n$ enumerates the states. $\Omega_\gamma$ ($\gamma=1, 2$) are dictated partly by the QR shape (see Fig. 1). Let us define as tunneling measure a parameter $\sigma = \dfrac{N_{n,1} - N_{n,2}}{N_{n,1} + N_{n,2}}$, with the range of [-1,1]. Obviously, when $\sigma = 0$, the electron will be located in $\Omega_1$ and $\Omega_2$ with equal probability. The case $|\sigma|\leq 1$ corresponds to the localization of the electron mainly in $\Omega_1$ or $\Omega_2$.

**Results of calculations**

Result of the calculations for the $\sigma$-parameter are presented in Fig. 3. The spectrum of a single electron can be classified by the bands which are related to the definite "radial quantum number" $n$. In Fig. 3 the bands are enumerated by $n=1, 2, 3, 4$. For each $n$-band the "orbital" quantum number $m$ is changing: $|m|=0,1,2,3,\ldots$. The set of the levels with fixed $n$ and increasing $|m|$ forms a typical shape as it is shown in Fig. 3. The shapes begin when $|m|=0$ and $\sigma$ is about 1. Increasing $|m|$ leads to decreasing $\sigma$ to a negative value of minimum, after that the increasing $|m|$ produces values for $\sigma$ which keep negative sign and approach to the zero asymptotically for large $|m|$. Each shape can be separated by the "beach" region and "whispering galleries" parts [20].

Let us remember that the quasi-doublets of the QR spectrum are described taking into account the double degeneracy of the eigenvalues of symmetric QR Hamiltonian of Eq. (1) for each value of the orbital quantum number $l$ (2). The degeneracy is lifted when the QR is asymmetrical ($\delta>0$). The obtained quasi-doublets are separated by the energy $\Delta E$.

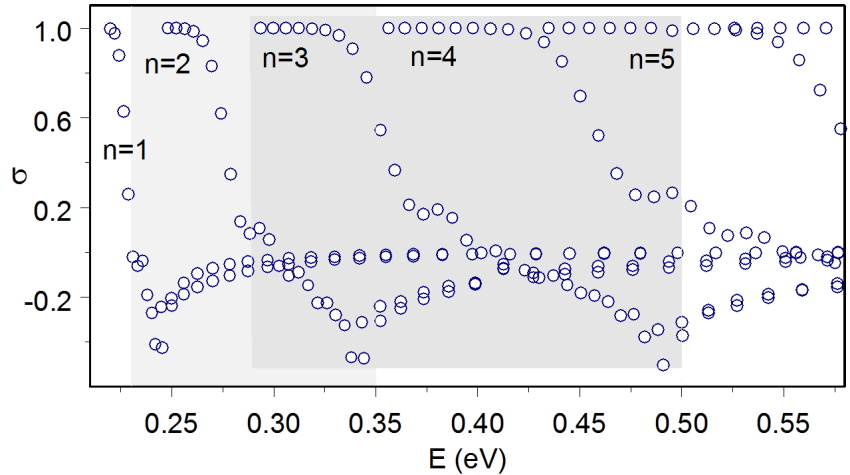

Figure 3. The parameter $\sigma$ and the single electron spectrum of the InAs/GaAs QR with $d = R-(r+\delta)=1.5(30-(19+4))$ nm. The bands are enumerated by $n=1, 2, 3, 4$. The shadowing energy intervals marks the "beach" regions for the $n=1$ and $n=2$ bands.

In the "beach" region (in Fig. 3 corresponding electron energies are within interval [0.225, 0.35] eV for the $n=1$ band), the splitting energies of the quasi-doublets decrease rapidly from visible values to zero. That agrees with experimental data [1] and previous analysis [20]. For the Bohigas billiard, "beach" region is defined by the relation $S \le r+\delta$ (with $R=1$) [20] which separates the chaotic region from regular one. For $S > r+\delta$ trajectories never hit the inner circle and motion is regular forming the so called "whispering galleries". In Fig. 3 the beach regions for the $n=1$ and $n=2$ bands are presented by shadowing of energy intervals. The quasi-doublets of the spectrum are well separated in the "beach" region and are visible as vertical oriented pairs. Wave function contour plots of quasi-doublet states existing in the "beach region" are presented in Fig. 4 for the $n=1$ band. The quasi-doublet wave functions are shifted in phase by the fixed angle. Note that variations of the QR geometry with fixed value of $(r+\delta)/R$ (with $r>\delta$) do not change this picture of the spectral dependence of the electron localization.

The energy spacing of the quasi-doublets and its dependence on confinement energy may be evidenced about chaotic properties of the QR [20]. In Ref. [1] it has been found an irregular behavior of $\Delta E/E$ as a function of $m/k$. The irregular peak was at the beginning of "whispering gallery" region of the $n=1$ band and was described in Ref. [1] as the effect of the "chaos assisted tunneling" due to the increase of the quasi-doublet splitting.

One can see from Fig. 3 that there is the irregular behavior of the $\sigma$ parameter just at right boundary of the "beach" region when the shapes of different $n$-band are crossed. It means that the wave functions are overlapped and distorted that is indicated by $\sigma$. We interpret it as results of anti-crossing of the levels of different $n$-bands. This anti-crossing affects on quasi-doublet splittings. This effect is indicated by the $\Delta E/E$ local enhancement as it was in experiments [1]. The anti-crossings occur for the levels with the same type angular symmetry (with the similar values of $|m|$) and different $n$.

Our interpretation of the experimental result of [1] is illustrated by Fig. 5. We decreased the QR sizes used in Fig. 3 up to 2 times. The less density spectrum gives clearer picture for the "beach region". Results of the calculations for the values of $\Delta E/E$ as a function of energy of the states are presented in Fig. 5. We have to note that the calculations are limited by the value of $\Delta E > 10^{-9}$ due to the accuracy of the numerical methods used.

The calculations reproduce qualitatively the experimental result of Ref. [1]. Fig. 5 shows results of our calculations of the $\sigma$-parameter and energy splitting of quasi doublets $\Delta E/E$ for different magnitudes of $\delta$. It has to be remarked that geometric configuration defined by the ratio $(r+\delta)/R$ is chosen close to the ratio 0.75 used in the paper [1]; we use it as 0.767. Increasing $\delta$ leads to the decreasing of the inner radius $r$. Comparing Fig. 5 a) and Fig. 5 d) we notice the displacement of energies of all levels of the spectrum by the values which are proportional to changing of $\delta$. Such behavior is natural because this leads to the increasing the "width" of the corresponding potential well. The spectrum is divided into bands with the numbers $n=1$, $n=2$, $n=3$. The data show a characteristic shapes with the intersections of the bands $n=1$ and $n=2$, which are displaced from 0.46 eV for Fig .5 a) to the 0.40 eV for Fig. 5 d), respectively. The intersections correspond to the situations when the wave functions of the bands with $n=1$ and $n=2$ have the similar spatial distributions and close energies. For the $n=1$ states there is an exponential increment of the quasi doublet splitting at the beginning of the "gallery" region. We interpret this effect as a result of the "level repulsion" at the anti-crossing. States of the different bands ($n=1$, $n=2$) with same type of the rotational symmetry are in the anti crossing situation. Their wave functions are overlapping. As is seen from Fig. 5, increment of the quasi doublet splitting depends on the parameter $\delta$, which determines the fact of the existence and magnitude of the splitting of quasi doublets in QR. Fig. 6 presents just the results of calculations of the maximal (over the total spectrum ) splitting $\Delta E/E$ as a function of the eccentricity $\delta$. Exponential dependence on Fig. 6 is in agreement with the general picture for tunneling in the quantum systems (see, for instance [20]).

It is worthy to remark that that the intersection of bands takes a place when the "beach" region of the $(n+1)$-band penetrates into the region of the "whispering galleries" [20] of the $n$-band. Indeed, in this region the magnitude of the perturbation of the quasi doublet splitting is maximal due to the anti crossing of the states having close values for the "orbital" quantum numbers even in this region. The "gallery" states of different $n$-bands are not anti-crossed, as well the states with difference of the "radial" quantum number larger than 1. Contour plot of the wave functions of quasi-doublet states of different radial quantum number bands ($n=1$ and $n=2$) are shown in Fig. 7. The anti-crossing occurs for upper member of the $n=2$ quasi-doublet and lower member of the $n=1$ quasi-doublet. Corresponding wave functions are distorted in radial direction with mixed symmetry. Other members of the doublets do not interact and the radial behavior of their wave functions corresponds strongly to "good" quantum number $n$. The same properties have wave functions of other states which are not involved in an anti-crossing (see Fig. 3 also).

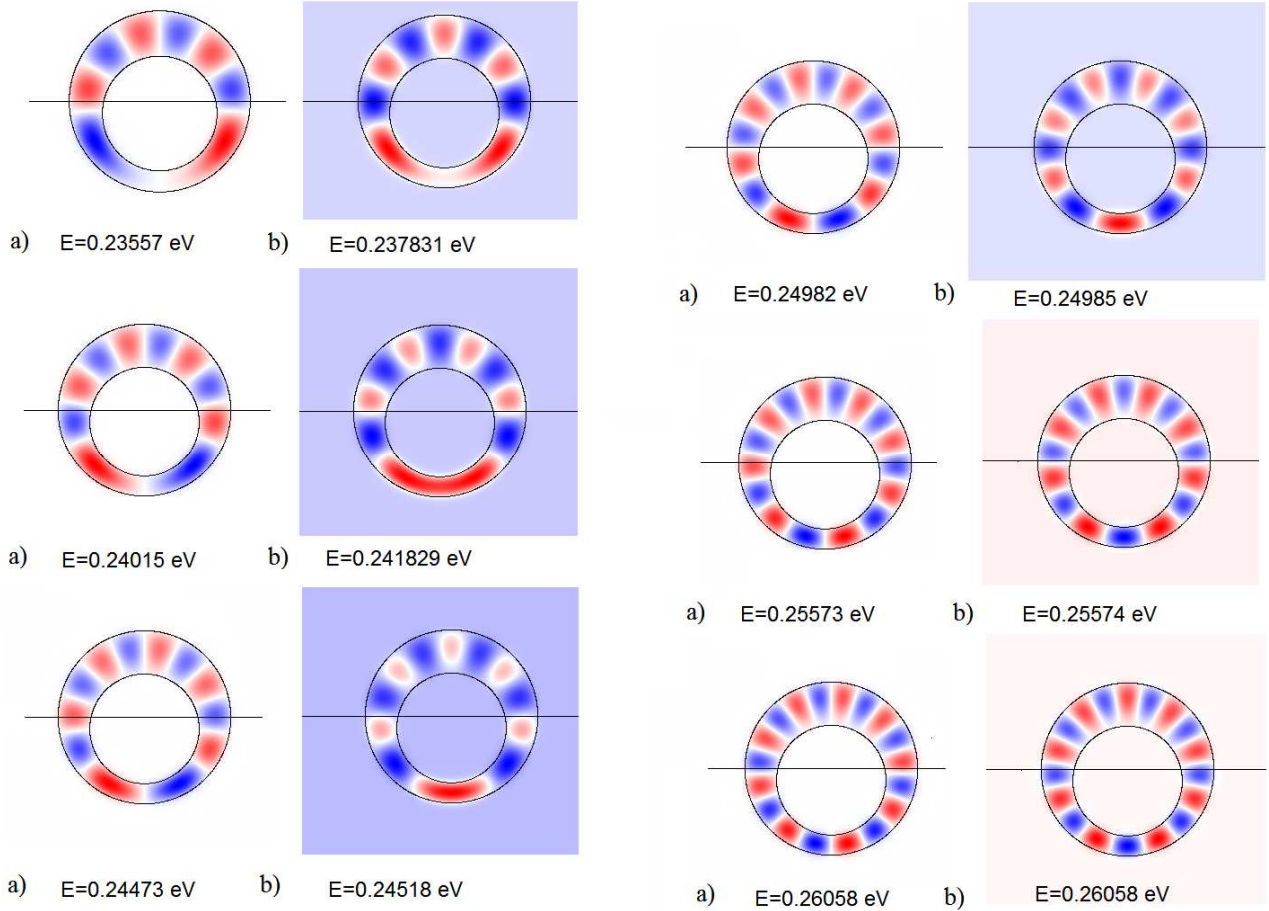

Figure 4. Single electron wave functions (contour plots) of quasi-doublet states: a) and b) for the lower and upper members of the quasi-doublets; (left) the states with quantum numbers $(n,|m|)=(1, \{8,10,12\})$; (right) $(n,|m|)=(1, \{14,16,18\})$. The electron energy of each state is shown. The QR geometry is given by $d=R-(r+\delta)=1.5(30-(19+4))$ nm.

It is known that the anti-crossing is accompanied by "level repulsion". In Fig. 7 we give an illustration for that. The anti-crossing level repulsion shown by green arrows leads to enhancement of the quasi-doublet splitting of the $n=1$ band, since the lower member of the doublet is involved into the repulsion. The degree of the distortion for wave functions of anti-crossed levels may be distinct for different quasi-doublets. Obviously, this difference correlates with the energy distance between the anti-crossed levels. In Fig. 7 and 8 one can see the different degrees of the distortion for wave functions of anti-crossed levels in the QRs with different geometry. The anti-crossing is a dynamical process. For the present consideration, the factor, which changes energies of the levels, is a variation of the geometry parameters. The change of the dynamical factor leads to the anti-crossing for the states with the same symmetry of wave function. Ideally, in the "crossing point" the wave functions of anti-crossed levels are the same. When the dynamical parameter is in small vicinity of this point, the wave functions differ. Thus the difference between the dynamical parameter in a given state and the one in "point of crossing" correlates with the degree of the distortion for wave functions of anti-crossed levels. Taking into account this statement, we see that in Fig. 8 the dynamical parameter is closer to the "point of crossing" because the distortion of the wave function is larger than in Fig. 7.

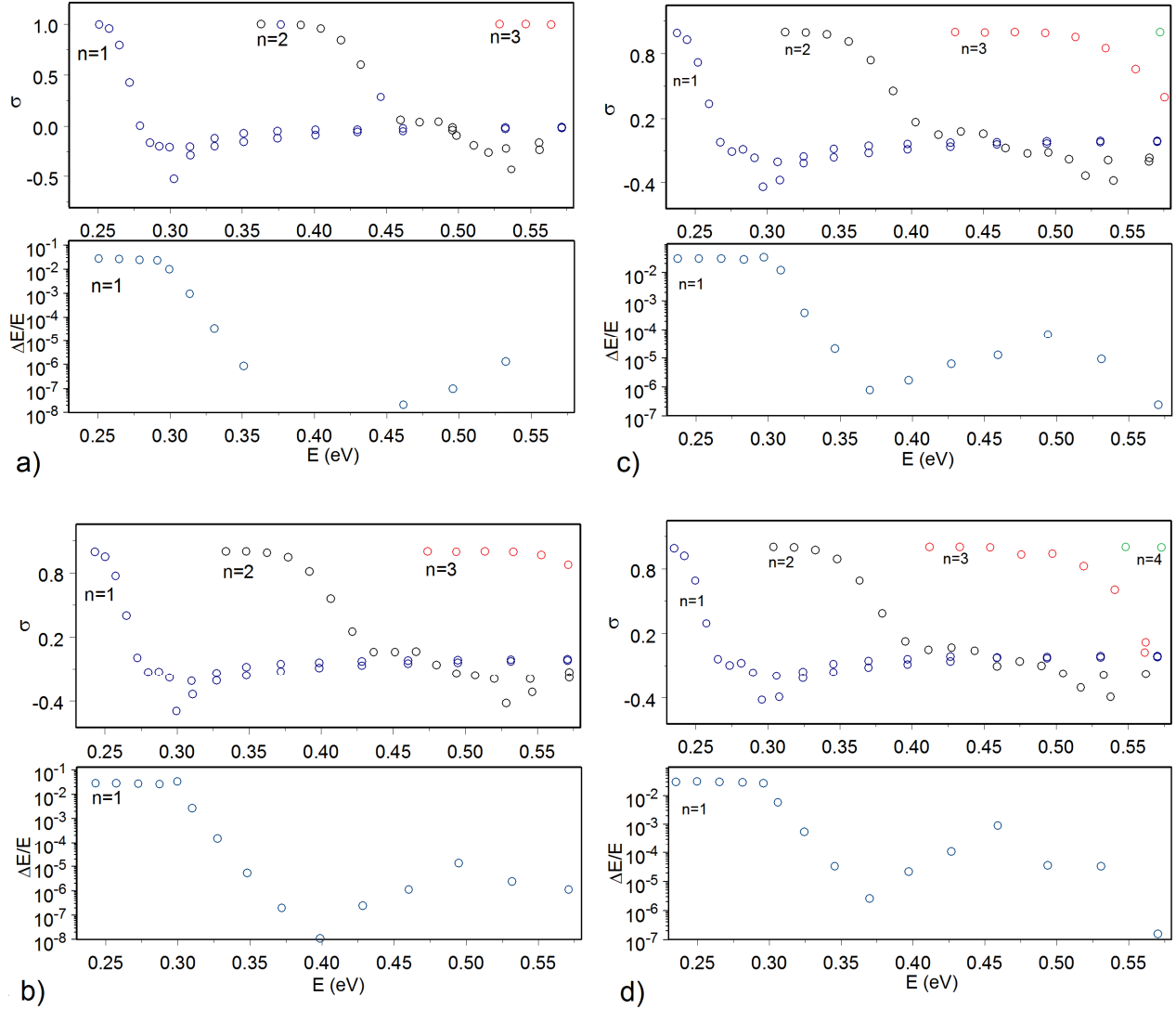

Figure 5. (Upper) The parameter $\sigma$ and the single electron spectrum of the InAs/GaAs QR with fixed value of $d=R-(r+\delta)=0.75(30-(20+3))$ nm and $S=(r+\delta)/R=0.767$ and different eccentricity $\delta$: a) $d=R-(r+\delta)=0.75(30-(20+3))$ nm; b) $d=R-(r+\delta)=0.75(30-(19+4))$ nm; c) $d=R-(r+\delta)=0.75(30-(18+5))$ nm; d) $d=R-(r+\delta)=0.75(30-(17.5+5.5))$ nm. The $n$-bands are numerated by $n=1$ (blue circles), $n=2$ (black circles), $n=3$ (red circles); (Lower) The energy of the relative spacing $\Delta E/E$ of the quasi-doublet for $n=1$ and different $|m|$.

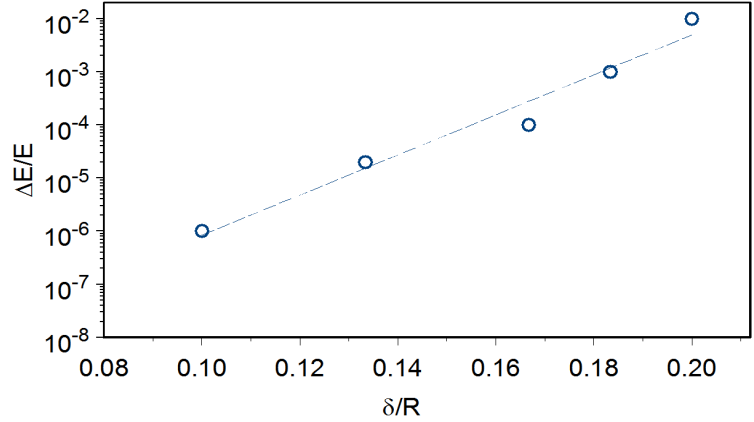

Figure 6. Maximal relative energy splitting $\Delta E/E$ (in logarithmical scale) of the $n=1$ band quasi-doublets in the electron spectrum for different values of the eccentricity $\delta$. The fixed geometry parameters of the InAs/GaAs QR are $R=30$ nm, $d=0.75(R-(r+\delta))$, $S=(r+\delta)/R=0.767$. The $r$ and $\delta$ are changed.

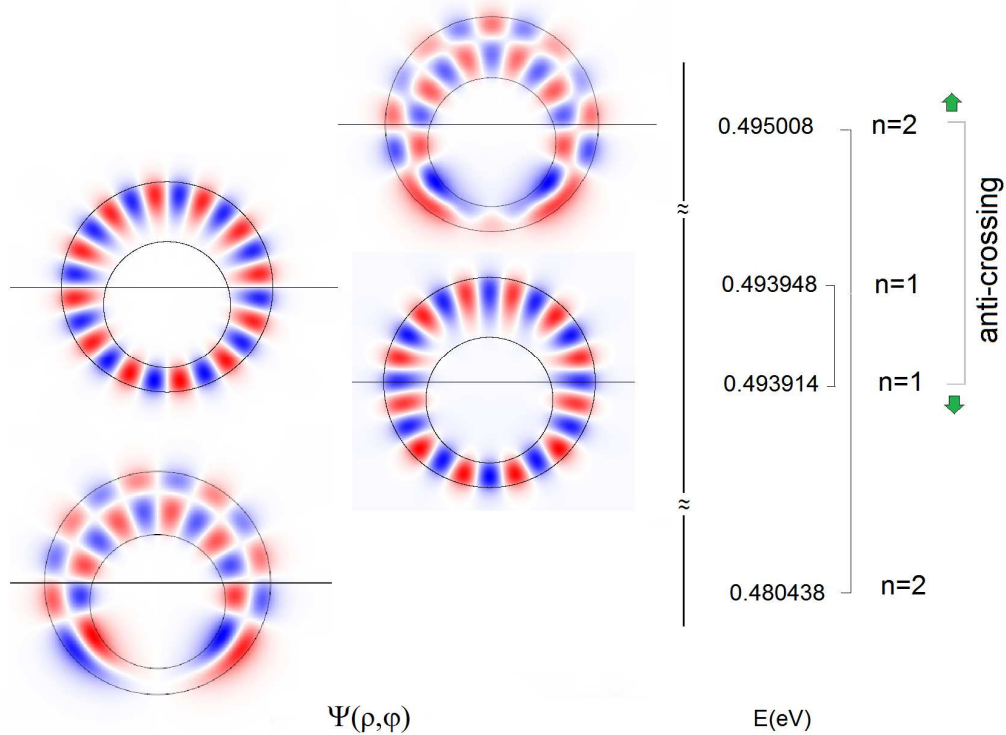

Figure 7. (Left) Contour plot of the wave functions of quasi-doublet states for different radial quantum number bands ($n=1$ and $n=2$). (Right) The energies of these states are shown. The distortion of the wave functions is detected for lower members of the doublets as a result of the anti-crossing of the levels. Geometry parameters are $d=R-(r+\delta)=0.75(30-(18+5))$ nm ($S=(r+\delta)/R=0.767$) (see Fig. 5). Energy gap between the anti-crossed levels is about $10^{-3}$ eV.

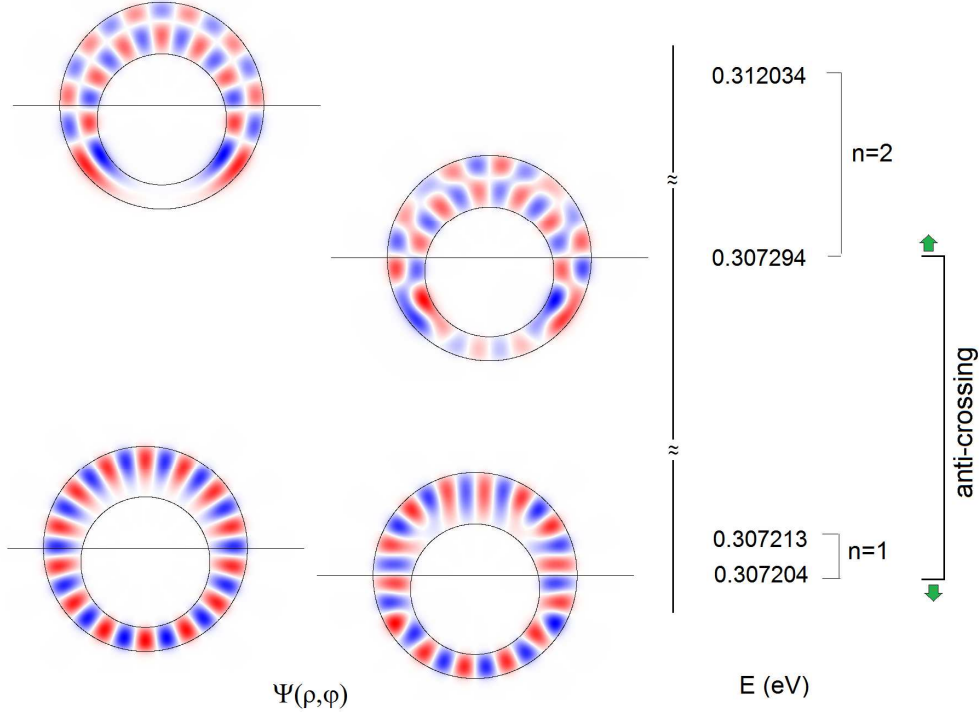

Figure 8. (Left) Contour plot of the wave functions of quasi-doublet states of different radial quantum number bands ($n$=1 and $n$=2). (Right) The energies of the states are shown. The notations are the same as in Fig. 7. Geometry parameters are $d=R-(r+\delta)=1.5(30-(18+5))$ nm ($S=(r+\delta)/R$=0.767). Energy gap between the anti-crossed levels is about $10^{-4}$ eV.

As it is clear from Fig. 5 and 6, the anti-crossing occurs between levels with the same rotation symmetry or, in other words, – same sign of "orbital" quantum number. We found that the inter-band tunneling occurs when the difference between the quantum number $|m|$ of the bands $n$=1 and $n$=2 is less them 5. In the opposite case the inter-band tunneling is forbidden. For this reason, only levels of the nearest $n$-bands ($n$ and $n$+1) are anti-crossed.

In Ref [20] the effect of the "chaos assisted tunneling" for the quantum billiard was expressed by random jumps of the tunneling rate $\Delta E/E$ when the eccentricity $\delta$ is regularly changed. The calculation have been performed for the states of the $n$=1 band with large $|m|/k$. In our consideration, this effect is explained by the inter-band anti-crossing under the condition of non-Poissonian distribution of the level spacing in the chaotic billiard. In other words, the regular change of the geometry at fixed $(r+\delta)/R$ leads to the corresponding change of the spectrum that means random change of anti-crossing position and the corresponding change of $\Delta E/E$. The existing of the chaos is essential for this effect.

Irregular case [20] corresponds to the QR geometry with $r<\delta$. The variables $(\rho,\varphi)$ in Eq. (1) are not separated. We have no "good" quantum numbers and our consideration based on the $\sigma$-parameters cannot single out the states with different quantum numbers $n$. An example of the calculations for the $\sigma$-parameter in the case $r<\delta$ is shown in Fig. 9. The characteristic shapes of the dependence of the $\sigma$-parameter on energy are hardly visible. We guess that all states are involved into an anti-crossing. The tunneled states (when $\sigma\approx 0$) which were dominated in the regular case of $r > \delta$ (small $\delta$) are in minority. The wave functions reflect the irregularity by the violation of up-down symmetries as it is shown in Fig. 10.

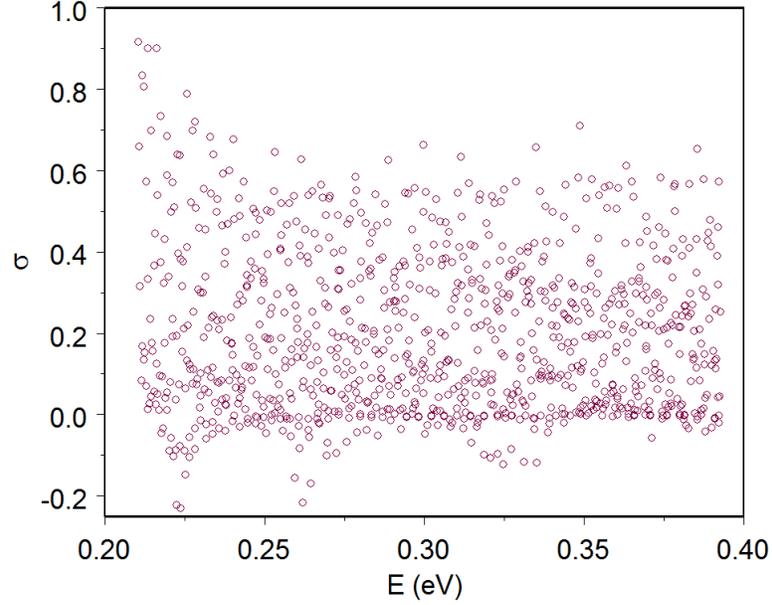

Figure 9. $\sigma$-parameter for InGaAs/GaAs QR with the geometry parameters $d = R-(r+\delta) = 3(40-(16+18))$ nm.

Such states dominate in the spectrum. We conclude that dynamical QR geometry transition from QR with small "chaos" ($\delta < r$) to QR with large "chaos" ($\delta > r$) is accompanied with the increasing number of the inter-band tunneling states due to anti-crossing.

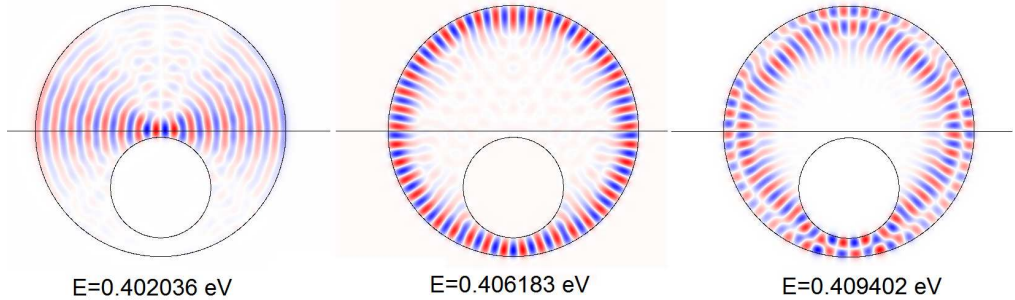

E=0.402036 eV      E=0.406183 eV      E=0.409402 eV

Figure 10. The contour plots of the wave functions of a single electron in the InAs/GaAs QR with $d = R-(r+\delta) = 60-(24+27)$ nm for several states. The energy of each state is shown.

**Conclusion**

In this paper we used $\sigma$-parameter to describe overall single electron localization in the chaotic InAs/GaAs QR. We found that, for small values of the QR eccentricity $\delta$, the spectrum of single electron confinement states can be classified by bands with "radial" quantum number $n$. Dependence of the $\sigma$-parameter on electron energy demonstrates characteristic shapes differ for each $n$. The shape behavior as a function of energy is alike for each $n$-band. The separation of the $n$-band on the "beach" and "gallery" parts is visible. The shapes of different bands are overlapped.

We found, that in the energy regions of the overlapping, the anti-crossing of the nearest levels occurred. The wave functions of these states are distorted relative to ones of non-anti-crossed states. The anti-crossing is accompanied by an energy gap between the energy levels involved. In the region of the "interaction" of the $n$ and ($n+1$)-bands, the energy splitting of quasi-doublets of $n$-band increase exponentially with increasing of eccentricity $\delta$. This increasing has been detected in the experiments of Ref. [1].

The anti-crossing provides the mechanism of the tunneling between the bands with different "radial quantum numbers". However this tunneling is not directly related to the chaotic properties of the considered objects. The analogous situation has a place for the double non-chaotic concentric QR in the perpendicular magnetic field [26]. The magnetic field lifting the degeneracy in the orbital quantum numbers provides conditions for anti-crossing of levels related with different rings, so the electron tunneling between the rings may be occurred. These levels have different radial and equal orbital quantum numbers (inter-band tunneling). The parameter $\delta$ for the considered here annular QR and the magnitude of the magnetic field govern the tunneling. The stochastic component of the chaotic QR results the random change of the QR spectrum when $\delta$ is changing. When the energies were fixed, the inter-band tunneling is determined exactly. Generally, violation of the confinement potential symmetry is the reason for anti-crossing and the inter-band tunneling for both these cases.

We can conclude that the stochastic properties of the QR electron spectrum do not play essential role for existing the inter-band tunneling. Thus, presented in Ref. [1] interpretation of the experimental data as the effect of the "chaos assisted tunneling", has the alternative, which we describe above.

For irregular case when inner radius $r$ of QR less than eccentricity $\delta$ ($r<\delta$) it is not possible separate $n$-bands by means the $\sigma$-parameter and we have seen that all states are involved into an anti-crossing. Dependence of the $\sigma$-parameter on energy is randomized in this case that reflects the chaotic properties of the QR. Anti-crossing is, of course, a level repulsion. When the amount of anti-crossed levels is high enough we have quantum chaos [27], which increases with increasing eccentricity $\delta$ and leads to extensive tunneling at large $\delta$ (see also [20]).